\documentclass[a4paper,11pt]{article}
\usepackage{jheppub}
\usepackage{placeins}
\usepackage[compat=1.1.0]{tikz-feynhand}

\tikzset{
	inflaton/.style={
		semithick,
		black,
		dash pattern=on 5pt off 3pt,
		line cap=round
	},
	axion/.style={
		semithick,
		black!65,
		dash pattern=on 2.5pt off 2pt,
		line cap=round
	}
}

\newcommand{\Mp}{M_{\rm P}}
\newcommand{\Trh}{T_{\rm rh}}

\newcommand{\DNeff}{\Delta N_{\rm eff}}
\newcommand{\Gphi}{\Gamma_\phi}
\newcommand{\Gaa}{\Gamma_{\phi\to aa}}
\newcommand{\grho}{g_*}
\newcommand{\gs}{g_{*,s}}
\newcommand{\Arh}{\mathcal A_{\rm rh}}
\newcommand{\Kann}{\mathcal K_{\rm ann}}
\newcommand{\cann}{c_{\rm ann}}
\newcommand{\af}{\mathfrak{a}_f}
\newcommand{\ai}{\mathfrak{a}_i}
\newcommand{\arh}{\mathfrak{a}_{\rm rh}}
\newcommand{\GeV}{\,{\rm GeV}}

\title{An EFT Map of Axion Dark Radiation from Reheating}

\author[]{Yong Xu}
\affiliation{McGill University Department of Physics \& Trottier Space Institute\\
	3600 Rue University, Montr\'eal, QC, H3A 2T8, Canada}
\emailAdd{yong.xu6@mcgill.ca}
\abstract{
	Light, weakly coupled sectors can retain information about the cosmological background in which they are produced. 
	We study light axions produced during reheating and their contribution to dark radiation, $\Delta N_{\rm eff}$. 
	We develop a shift-symmetric EFT in which an inflaton-dependent axion kinetic term systematically organizes the leading production channels. 
	The same kinetic function generates both direct inflaton decay and inflaton annihilation from the oscillating inflaton background. 
	Direct decay is described by an invisible inflaton branching fraction, while annihilation is a genuinely reheating-sensitive source controlled by a coherent combination of Wilson coefficients.
	We derive the contribution to $\Delta N_{\rm eff}$ from both channels and show that they scale oppositely with the reheating temperature: the decay contribution falls as $T_{\rm rh}^{-2}$, whereas the annihilation contribution grows approximately as $T_{\rm rh}^{4/3}$. 
	Their crossing is missed by treatments that keep only one production channel. 
	We translate current and projected $\Delta N_{\rm eff}$ sensitivities into constraints on the Wilson coefficients of the kinetic function, obtaining a two-dimensional EFT map of axion dark radiation from reheating. 
	This map can imply both lower and upper bounds on the reheating temperature, showing that light axion relics can turn dark radiation measurements into constraints on reheating.
}

\begin{document}
	\begin{flushright}
		May  2026
	\end{flushright}
	\maketitle
	\section{Introduction}
	\label{sec:intro}
	
	Cosmic inflation remains one of the leading paradigms for resolving major puzzles of the early Universe, including the horizon problem, the flatness problem, and the origin of cosmic structure~\cite{Starobinsky:1980te,Guth:1980zm,Linde:1981mu,Albrecht:1982wi}.   In its simplest realizations, inflation is driven by a scalar field slowly rolling along a sufficiently flat potential.  After inflation ends, the inflaton reaches the vicinity of the potential minimum and begins to oscillate.  The energy stored in this oscillating inflaton background must then be transferred to lighter degrees of freedom, either in the Standard Model or in hidden sectors.  This process populates the Universe with relativistic particles and eventually produces the thermal bath of the hot Big Bang.  The transition from the inflaton-dominated stage to the radiation-dominated Universe is known as reheating~\cite{Allahverdi:2010xz,Amin:2014eta,Lozanov:2019jxc,Barman:2025lvk,Brandenberger:2026trv}.
	
	Although reheating fixes the earliest thermal history accessible to standard cosmology, its microscopic origin remains largely unknown.  The challenge is observational: the visible particles produced during reheating rapidly thermalize, and this thermalization erases much of the information about the operator that drained the inflaton energy. A useful reheating probe should therefore be weakly coupled enough to avoid thermalization, so that the information it carries is not erased.
	
	This makes weakly coupled relics particularly valuable.  A particle produced during reheating but never brought into equilibrium can preserve information that the visible bath loses.  This logic has been explored for weakly coupled particles such as FIMPs~\cite{Becker:2023tvd,Barman:2024nhr}, gravitons~\cite{Nakayama:2018ptw,Ema:2020ggo,Barman:2023ymn,Barman:2023rpg,Bernal:2023wus,Choi:2024ilx,Barman:2024htg,Xu:2024fjl,Xu:2025wjq,Wang:2026ule}, millicharged particles~\cite{Gan:2023jbs}, dark photons~\cite{Bernal:2024ndy,ShamsEsHaghi:2025kci}, axions and axion-like particles (ALPs)~\cite{Xu:2023lxw}, and sterile neutrinos~\cite{Cline:2026cek}.
	
	In this work we focus on axions and ALPs produced during reheating.  One well-studied possibility is that the axion becomes cold dark matter~\cite{Grin:2007yg, Visinelli:2009kt,Kobayashi:2020ryx,Blinov:2019jqc,Arias:2023wyg,Xu:2023lxw,OHare:2024nmr}.  During reheating, the expansion rate and entropy production differ from the standard radiation-dominated history, and this can modify the onset of axion oscillations and the final relic abundance.  In particular, Ref.~\cite{Xu:2023lxw} studied QCD axion and ALP dark matter from misalignment during inflationary reheating and showed that reheating can reshape the viable parameter space.  Here we take the complementary limit: the axion remains light and relativistic, so the imprint of reheating appears not as cold dark matter but as dark radiation, quantified by $\Delta N_{\rm eff}$.

	Several pieces of this physics have already appeared in the literature.
	Dark radiation from the decay of heavy nonrelativistic fields has been studied as a probe of the decay history and of changes in the relativistic degrees of freedom~\cite{Jaeckel:2021gah}.
	Axion dark radiation from modulus decays is well known in string-motivated reheating scenarios~\cite{Cicoli:2012aq,Higaki:2012ar}, while axion-like dark radiation from modulus or inflaton decays has also been studied phenomenologically~\cite{Jaeckel:2021ert,Ghoshal:2023phi}.
	Related axion kinetic functions have been discussed in the axiverse context~\cite{Gorbunov:2017ayg}.
	Axion production from inflaton scattering or kinetic couplings has also been studied in more model-dependent settings~\cite{Lee:2023dtw}, and  pNGB $\chi$ production from the derivative operators $\phi(\partial\chi)^2$ and $\phi^2(\partial\chi)^2$ during reheating has been analyzed using the coherent oscillating inflaton background, including an application to Peccei--Quinn inflation with a large non-minimal coupling~\cite{Kaneta:2024yyn}.
	
	These developments show that axions are natural carriers of reheating information.
	However, they also reveal a limitation of treating individual production mechanisms separately.
	In the direct-decay approach, the result is reduced to an invisible branching fraction.
	In scattering or kinetic-production studies, the calculation is often tied to a specific model, or to derivative operators treated as separate production sources.
	For a shift-symmetric axion, this is not the most natural organization.
	The leading couplings to the inflaton arise from the inflaton dependence of the axion kinetic metric.
	In other words, the object to expand is not a single operator, but the function multiplying $(\partial a)^2$ with $a$ being the axion.
	
	This observation motivates the EFT framework of this paper.  We write the axion kinetic term as an inflaton-dependent metric, $Z(\phi)(\partial a)^2/2$, and expand $Z(\phi)$ near the minimum of the inflaton potential.  The linear response of $Z(\phi)$ generates direct inflaton decay, $\phi\to aa$.  The quadratic response generates a contact contribution to inflaton annihilation, $\phi\phi\to aa$.  But the annihilation amplitude is not determined by the quadratic coefficient alone: two insertions of the linear response also generate exchange diagrams.  Thus a consistent kinetic EFT automatically ties together decay and annihilation, and the two-inflaton channel is controlled by the coherent amplitude of the full kinetic function rather than by $\phi^2(\partial a)^2$ alone.
	
	The goal is therefore not merely to compute another dark-radiation abundance.  The goal is to identify what information $\DNeff$ carries about the microscopic reheating operator when the axion coupling is organized by symmetry.  In the EFT map developed below, decay-produced dark radiation measures the invisible branching fraction of the inflaton, while annihilation-produced dark radiation measures the reheating history of the oscillating inflaton background.  These two sources have opposite dependence on the reheating temperature: direct decay is strongest at low $T_{\rm rh}$, whereas inflaton annihilation can become important at high $T_{\rm rh}$.  Their competition produces phenomenological features, including possible lower and upper bounds on $T_{\rm rh}$, that are absent in decay-only or annihilation-only descriptions.
	
	The paper is organized as follows.  Section~\ref{sec:setup} defines the reheating convention, the axion kinetic EFT, and a possible UV origin.  Section~\ref{sec:DR} presents the general conversion from a decoupled axion radiation fraction to $\Delta N_{\rm eff}$.  Section~\ref{sec:decay} derives the direct-decay contribution and its interpretation as an invisible-branching-ratio bound.  Section~\ref{sec:annihilation} derives the inflaton-annihilation contribution, including the coherent Wilson-coefficient combination and the reheating-history kernel.  Section~\ref{sec:map} combines the two effects into the EFT map and shows how current and projected $\Delta N_{\rm eff}$ sensitivities constrain the $(c_1,c_2)$ plane.  We summarize the main findings in Section~\ref{sec:conclusions}.  The appendix provides the analytic approximation to the annihilation kernel.
	
	\section{Setup}
	\label{sec:setup}
	
	This section collects the assumptions and conventions used throughout the paper. We first describe the perturbative reheating background, then introduce the axion kinetic EFT, and finally discuss a simple UV origin.
	
	\subsection{Reheating background}
	After inflation, the inflaton starts oscillating around the potential minimum. Here, we assume the inflaton oscillates in a quadratic potential,
	\begin{equation}
		V(\phi)=\frac12m_\phi^2\phi^2 .
		\label{eq:quadratic_potential}
	\end{equation}
	where $m_\phi$ denotes the inflaton mass around the minimum. A quadratic minimum of the form Eq.~\eqref{eq:quadratic_potential} is realized in many viable inflation models, including  Starobinsky\footnote{Recent results from the Atacama Cosmology Telescope (ACT) have been interpreted as placing the Starobinsky model under tension with observational data at more than the $2\sigma$ level~\cite{AtacamaCosmologyTelescope:2025nti}. However, the conclusion in the experimental paper \cite{AtacamaCosmologyTelescope:2025nti}  is based on a comparison employing only the leading-order approximation of the Starobinsky inflationary predictions. When higher-order corrections are consistently included, the theoretical predictions are shifted in a way that restores agreement with the ACT data at the $2\sigma$ level~\cite{Drees:2025ngb}. A careful and consistent comparison between theoretical predictions and observational data is therefore essential, particularly when higher-order effects lead to non-negligible shifts in the predicted observables. } inflation~\cite{Starobinsky:1980te}, attractor inflation~\cite{Kallosh:2013hoa}, and polynomial inflation~\cite{Drees:2022aea}.   Averaged over many oscillations, the inflaton behaves as pressureless matter. We denote by $\rho_\phi$ the averaged inflaton energy density and by $\rho_R$ the energy density in the visible radiation bath. The Hubble rate is $H=\dot{\mathfrak{a}}/\mathfrak{a}$, where $\mathfrak{a}$ is the scale factor.  The visible reheating process is parameterized by a decay width $\Gphi$. The background equations are
	\begin{align}
		\dot\rho_\phi+3H\rho_\phi&=-\Gphi\rho_\phi,
		\label{eq:bg_phi}\\
		\dot\rho_R+4H\rho_R&=\Gphi\rho_\phi,
		\label{eq:bg_R}\\
		H^2&=\frac{\rho_\phi+\rho_R}{3\Mp^2}.
		\label{eq:bg_H}
	\end{align}
	Here $\Mp=2.435\times10^{18}\GeV$ is the reduced Planck mass. The axion energy density is treated as a perturbative probe and is therefore not included in Eq.~\eqref{eq:bg_H}. This is consistent as long as the final dark-radiation fraction is small compared with unity.
	
	We use the reheating-temperature convention in which the end of reheating is estimated by
	\begin{equation}
		H(\Trh) = H(\arh)\simeq \frac{2}{3}\Gphi\,,
		\label{eq:H_Trh_convention}
	\end{equation}
	where $\Trh$ denotes the temperature when  the scale factor $a= \arh$.
	This convention follows the matter-dominated intuition $H\simeq2/(3t)$ and identifies the end of reheating with the epoch when the microscopic decay time becomes comparable to the Hubble time. The visible radiation energy density at temperature $T$ is $\rho_R=(\pi^2/30)\grho(T)T^4$, where $\grho(T)$ is the effective number of relativistic energy degrees of freedom. Using Eq.~\eqref{eq:H_Trh_convention}, the reheating temperature is \cite{Drees:2021wgd}
	\begin{equation}
		\Trh
		=
		\sqrt{\frac{2}{\pi}}
		\left(\frac{10}{\grho(\Trh)}\right)^{1/4}
		\sqrt{\Mp\Gphi}.
		\label{eq:Trh_def}
	\end{equation}
	Equivalently,
	\begin{equation}
		\Gphi
		=
		\frac{\pi}{2}
		\left(\frac{\grho(\Trh)}{10}\right)^{1/2}
		\frac{\Trh^2}{\Mp}.
		\label{eq:Gamma_Trh}
	\end{equation}
	All formulae below use this convention. 
	
	\subsection{Inflaton-dependent axion kinetic term}
	\begin{figure}[t]
		\centering
		
		\begin{tikzpicture}
			\begin{feynhand}
				
				\vertex (p) at (-2.0,0) {$\phi$};
				\vertex [dot] (v) at (0,0) {};
				\vertex (a1) at (1.7,0.8) {$a$};
				\vertex (a2) at (1.7,-0.8) {$a$};
				
				\propag [inflaton] (p) to (v);
				\propag [axion] (v) to (a1);
				\propag [axion] (v) to (a2);
				
				\node at (0,1.35) {(a) decay};
				\node at (0,-1.25) {$c_1$};
				
				\vertex (p1) at (4.0,0.8) {$\phi$};
				\vertex (p2) at (4.0,-0.8) {$\phi$};
				\vertex [dot] (vc) at (5.8,0) {};
				\vertex (b1) at (7.6,0.8) {$a$};
				\vertex (b2) at (7.6,-0.8) {$a$};
				
				\propag [inflaton] (p1) to (vc);
				\propag [inflaton] (p2) to (vc);
				\propag [axion] (vc) to (b1);
				\propag [axion] (vc) to (b2);
				
				\node at (5.8,1.35) {(b) contact};
				\node at (5.8,-1.25) {$c_2$};
				
			\end{feynhand}
		\end{tikzpicture}
		
		\vspace{0.8cm}
		
		\begin{tikzpicture}
			\begin{feynhand}
				
				\vertex (p1) at (-2.0,0.9) {$\phi$};
				\vertex (p2) at (-2.0,-0.9) {$\phi$};
				
				\vertex [dot] (vtop) at (0,0.65) {};
				\vertex [dot] (vbot) at (0,-0.65) {};
				
				\vertex (a1) at (2.0,0.9) {$a$};
				\vertex (a2) at (2.0,-0.9) {$a$};
				
				\propag [inflaton] (p1) to (vtop);
				\propag [inflaton] (p2) to (vbot);
				\propag [axion] (vtop) to (vbot);
				\propag [axion] (vtop) to (a1);
				\propag [axion] (vbot) to (a2);
				
				\node at (0,1.55) {(c) $t$ channel};
				\node at (0,-1.55) {$c_1^2$};
				
				\vertex (q1) at (4.0,0.9) {$\phi$};
				\vertex (q2) at (4.0,-0.9) {$\phi$};
				
				\vertex [dot] (utop) at (6.0,0.65) {};
				\vertex [dot] (ubot) at (6.0,-0.65) {};
				
				\vertex (c1) at (8.0,0.9) {$a$};
				\vertex (c2) at (8.0,-0.9) {$a$};
				
				\propag [inflaton] (q1) to (utop);
				\propag [inflaton] (q2) to (ubot);
				\propag [axion] (utop) to (ubot);
				\propag [axion] (utop) to (c2);
				\propag [axion] (ubot) to (c1);
				
				\node at (6.0,1.55) {(d) $u$ channel};
				\node at (6.0,-1.55) {$c_1^2$};
				
			\end{feynhand}
		\end{tikzpicture}
		
		\caption{
			Tree-level diagrams generated by the inflaton-dependent axion kinetic term.
			The linear operator $\phi(\partial a)^2$ gives the decay diagram (a), but two insertions of the same operator also contribute to $\phi\phi\to aa$ through the exchange diagrams (c) and (d).
			The quadratic operator $\phi^2(\partial a)^2$ gives the contact diagram (b).
			Therefore the annihilation amplitude is the coherent sum of (b)--(d), rather than the contact diagram alone.
			This is why the nonrelativistic annihilation amplitude depends on the combination $c_2-c_1^2$.
		}
		\label{fig:feynman_diagrams}
	\end{figure}
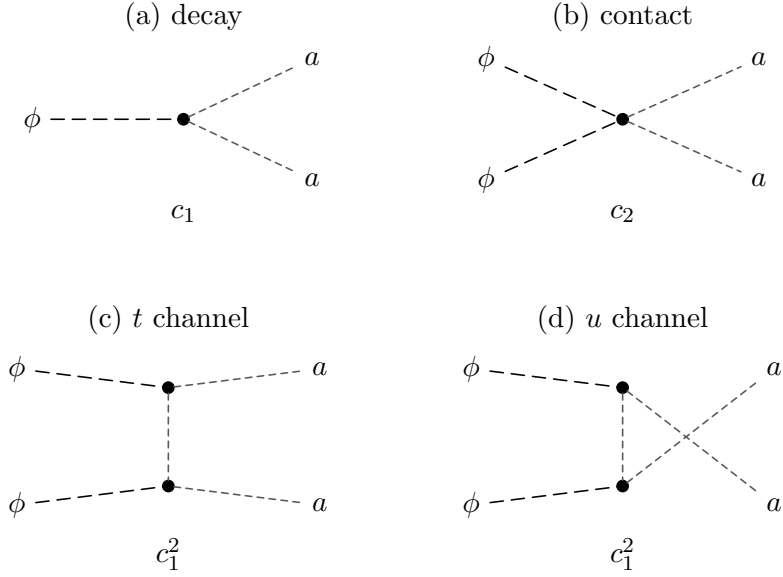
	
	We consider a light axion-like field $a$ whose interactions respect the perturbative shift symmetry
	$a\to a+{\rm constant}$.  The axion is assumed to remain out of equilibrium with the visible bath after production.  The leading model-independent way in which the inflaton can communicate with such a shift-symmetric axion is through the axion kinetic metric.  We therefore write
	\begin{equation}
		\mathcal L
		\supset
		\frac12(\partial\phi)^2
		-
		\frac12m_\phi^2\phi^2
		+
		\frac12 Z(\phi)\,\partial_\mu a\,\partial^\mu a
		+
		\mathcal L_{\rm rh}.
		\label{eq:Lagrangian}
	\end{equation}
	Here $\mathcal L_{\rm rh}$ denotes the unspecified visible-sector reheating interaction responsible for the width $\Gphi$.  The function $Z(\phi)$ specifies how the axion normalization responds to the inflaton background.  This is the object we want to map onto the observable dark-radiation abundance.
	
	Near the minimum of the inflaton potential, the most general local expansion of the kinetic metric is
	\begin{equation}
		Z(\phi)
		=
		1
		+
		c_1\frac{\phi}{\Lambda}
		+
		\frac{c_2}{2}\frac{\phi^2}{\Lambda^2}
		+
		{\cal O}\left(\frac{\phi^3}{\Lambda^3}\right),
		\label{eq:Z_expansion}
	\end{equation}
	where $\Lambda$ is the scale suppressing the inflaton dependence of the axion kinetic term, and $c_1,c_2$ are dimensionless Wilson coefficients.  We keep $\Lambda$ explicit throughout the analysis.  The choice $\Lambda=M_P$ is a useful benchmark for Planck-suppressed kinetic couplings, but it is not assumed.  Smaller values of $\Lambda$ describe stronger inflaton--axion kinetic interactions and enhance the production rates.  This enhancement, however, must be accompanied by a controlled expansion of $Z(\phi)$: the perturbative description is reliable only over the range of inflaton oscillation amplitudes for which the corrections to the axion kinetic metric remain small and $Z(\phi)$ stays positive.
	
	This point is important for lower-cutoff benchmarks.  If $\Lambda\ll M_P$, the expansion of $Z(\phi)$ need not be valid immediately after inflation, when the inflaton amplitude can be Planckian.  In that case the perturbative EFT calculation should be initialized only after the oscillation amplitude has redshifted into the domain where Eq.~\eqref{eq:Z_expansion} is under control.  We denote this amplitude by $\Phi_i$.  A useful way to parameterize a controlled onset is
	\begin{equation}
		\Phi_i=\xi\Lambda,
		\qquad
		\xi<1,
		\label{eq:Phi_xi}
	\end{equation}
	with $\xi$ chosen small enough that the kinetic metric is weakly modulated.  This does not assume that reheating begins at $\Phi_i$; it specifies the onset of the perturbative kinetic-EFT description used to compute the particle-level decay and annihilation contributions.  The earlier epoch, if $\Phi\gtrsim\Lambda$, may require a UV completion of the kinetic function.
	
	The interactions relevant at this order are
	\begin{equation}
		\mathcal L_{\rm int}
		=
		\frac{c_1}{2\Lambda}\phi\,\partial_\mu a\partial^\mu a
		+
		\frac{c_2}{4\Lambda^2}\phi^2\,\partial_\mu a\partial^\mu a .
		\label{eq:interactions}
	\end{equation}
	The two terms in Eq.~\eqref{eq:interactions} describe physically distinct ways of transferring inflaton energy into axions.  The linear operator mediates direct decay, $\phi\to aa$.  The quadratic operator gives a contact contribution to $\phi\phi\to aa$.  However, the two-inflaton amplitude is not determined by the quadratic operator alone.  Two insertions of the linear operator also generate axion exchange diagrams; see Fig.~\ref{fig:feynman_diagrams}.  Therefore a consistent EFT treatment must keep the linear and quadratic terms together.
	
	For later convenience we define
	\begin{equation}
		\cann
		\equiv
		c_2-c_1^2 .
		\label{eq:cann_def}
	\end{equation}
	As  shown in Sec.~\ref{sec:annihilation}, this is the combination that controls $\phi\phi\to aa$ in the nonrelativistic limit relevant for the oscillating inflaton.  It arises because the contact diagram from $\phi^2(\partial a)^2$ must be combined coherently with the exchange diagrams generated by two insertions of $\phi(\partial a)^2$.  The annihilation channel is therefore a probe of the full kinetic metric, not of $c_2$ in isolation.
	
	A related kinetic-function setup was considered in Ref.~\cite{Gorbunov:2017ayg} in the context of axiverse dark radiation.  In that work the abundance was estimated from the linear decay term.  Here we keep the two leading terms in the same EFT expansion and compute both direct decay and inflaton annihilation.  This is essential for identifying the physical annihilation combination $c_2-c_1^2$, which is the quantity constrained by $\DNeff$ below.
	
	\subsection{A possible UV origin}
	We now describe a simple possible UV origin of the operators in Eq.~\eqref{eq:interactions}.  Consider a Peccei--Quinn sector with complex PQ field $P$.  Below the PQ-breaking scale we may write
	\begin{equation}
		P(x)
		=
		\frac{1}{\sqrt2}\,[F+s(x)]\,e^{ia(x)/F},
	\end{equation}
	where $F$ is the PQ-breaking scale and $s$ is the radial mode, or saxion.  The PQ kinetic term contains
	\begin{align}
		|\partial P|^2
		&\supset
		\frac12\left(1+\frac{s}{F}\right)^2
		\partial_\mu a\,\partial^\mu a
		\nonumber\\
		&=
		\frac12\partial_\mu a\,\partial^\mu a
		+
		\frac{s}{F}\partial_\mu a\,\partial^\mu a
		+
		\frac{s^2}{2F^2}\partial_\mu a\,\partial^\mu a .
		\label{eq:PQ_kinetic_origin}
	\end{align}
	The axion kinetic term is therefore controlled by the saxion background.  Portal interactions between the inflaton and the PQ sector,
	\begin{equation}
		\mathcal L
		\supset
		\mu\,\phi |P|^2
		+
		\lambda\,\phi^2 |P|^2 ,
		\label{eq:PQ_portals}
	\end{equation}
	then generate an inflaton-dependent axion kinetic metric after the saxion is integrated out.  The linear and quadratic portals generate the corresponding linear and quadratic responses in the low-energy kinetic function.  When the portal interactions contain terms linear in the saxion after PQ breaking, these operators arise from tree-level threshold matching through saxion exchange.  More generally, the same operator structures are generated radiatively by loop diagrams with virtual saxions.  The coefficients $c_1$ and $c_2$ in Eq.~\eqref{eq:interactions} should therefore be understood as low-energy Wilson coefficients encoding the response of the PQ radial sector to the inflaton background.
	
	This UV picture also makes clear why the decay and annihilation channels can be separated by symmetries.  If a $\phi\to-\phi$ symmetry forbids the linear portal, then the operator $\phi(\partial a)^2$ is absent, while $\phi^2(\partial a)^2$ remains allowed.  Inflaton annihilation can then be the leading perturbative axion-production channel.  In the EFT analysis below we keep both operators, allowing for the most general leading response of the axion kinetic metric.
	\section{Dark-radiation dictionary}
	\label{sec:DR}
	In this section we relate the axion radiation fraction produced during reheating to $\Delta N_{\rm eff}$.
	To this end, we define
	\begin{equation}
		R_a
		\equiv
		\frac{\rho_a}{\rho_R}\bigg|_{\Trh},
		\label{eq:R_def}
	\end{equation}
	where $\rho_R$ is the visible radiation energy density and $\rho_a$ is the energy density stored in relativistic axions.  The ratio is evaluated after visible reheating has completed with $\mathfrak{a} = \arh$ or $T = \Trh$.  After this time both sectors redshift as radiation, but only the visible sector is heated by the entropy release from particles that later become nonrelativistic.  This entropy release is the only effect needed to convert $R_a$ into $\DNeff$.
	
	The axions are decoupled, so their temperature redshifts simply as $T_a\propto \mathfrak{a}^{-1}$.  By contrast, the visible bath obeys entropy conservation, $\gs(T)\,T^3\mathfrak{a}^3={\rm const.}$
	Between reheating and neutrino decoupling this gives
	\begin{equation}
		\frac{\arh \Trh}{\mathfrak{a}_{\nu{\rm dec}} T_{\nu{\rm dec}}}
		=
		\left[
		\frac{\gs(T_{\nu{\rm dec}})}{\gs(\Trh)}
		\right]^{1/3}.
		\label{eq:Ta_Tnu}
	\end{equation}
	Thus the decoupled axion bath is colder than the visible bath at neutrino decoupling.  Since the visible radiation density is
	$\rho_R=(\pi^2/30)\grho T^4$, the axion-to-visible radiation ratio at neutrino decoupling is
	\begin{align}
		\left.
		\frac{\rho_a}{\rho_R}
		\right|_{T_{\nu{\rm dec}}}
		&= \left[ \rho_a(\Trh) \left(\frac{\arh}{\mathfrak{a}_{\nu{\rm dec}}}\right)^4 \right] \Bigg / \left[\rho_R(\Trh)  \frac{\grho (T_{\nu{\rm dec}})}{\grho(\Trh)} \left(\frac{T_{\nu{\rm dec}}}{\Trh}\right)^4 \right]\nonumber \\
		&  = 
		R_a\,
		\frac{\grho(\Trh)}{\grho(T_{\nu{\rm dec}})}
		\left[
		\frac{\gs(T_{\nu{\rm dec}})}{\gs(\Trh)}
		\right]^{4/3}.
		\label{eq:R_nudec}
	\end{align}
	The factor involving $\grho$ accounts for the change in the visible radiation density, while the entropy factor accounts for the relative cooling of the decoupled axions.
	
	At neutrino decoupling we take $\grho(T_{\nu{\rm dec}})=\gs(T_{\nu{\rm dec}})=10.75$.
	One fully thermalized neutrino species then carries the fraction $\rho_{\nu,1}/\rho_R
	=\frac{7/4}{10.75}$
	of the total visible radiation density.  Therefore
	\begin{equation}
		\DNeff
		\equiv
		\frac{\rho_a}{\rho_{\nu,1}}
		=
		\Arh R_a,
		\label{eq:DNeff_R}
	\end{equation}
	where
	\begin{equation}
		\Arh
		=
		\frac{4}{7}\grho(\Trh)
		\left[
		\frac{10.75}{\gs(\Trh)}
		\right]^{4/3}.
		\label{eq:Arh}
	\end{equation}
	For high reheating temperature, 
	$\grho(\Trh)=\gs(\Trh)=106.75$, giving
	\begin{equation}
		\Arh\simeq2.86 .
		\label{eq:Arh_SM}
	\end{equation}
	
	At leading order in the axion abundance, the final radiation fraction is the sum of the direct-decay and inflaton-annihilation contributions,
	\begin{equation}
		R_a
		=
		R_a^{\rm dec}
		+
		R_a^{\rm ann}.
		\label{eq:R_sum}
	\end{equation}
	Combining the dark-radiation dictionary with the perturbative production rates derived below gives the central EFT map,
	\begin{equation}
		\DNeff(c_1,c_2)
		=
		\Arh
		\left[
		{\rm BR}(\phi\to aa) 
		+
		\cann^2 \times \Kann
		\right]\,.
		\label{eq:master_formula}
	\end{equation}
	The first term is the direct-decay contribution, controlled by the invisible inflaton branching ratio ${\rm BR}(\phi\to aa)$, and we will derive it in  Sec.~\ref{sec:decay}.   The second term is the inflaton-annihilation contribution. It probes the coherent two-inflaton response of the kinetic EFT, with $c_{\rm ann}=c_2-c_1^2$, while $ \Kann$ is the reheating-history kernel derived in Sec. 5. Eq.~\eqref{eq:master_formula} is the main formula used throughout the rest of the paper.
	\section{Direct decay}
	\label{sec:decay}
	
	The direct-decay channel is the simplest limit of the EFT map.   The linear kinetic operator in Eq.~\eqref{eq:interactions} allows a single inflaton quantum to decay into two axions, $\phi\to aa$.
	For $m_a\ll m_\phi$, the decay width is
	\begin{equation}
		\Gaa
		=
		\frac{c_1^2}{128\pi}
		\frac{m_\phi^3}{\Lambda^2},
		\label{eq:Gamma_aa}
	\end{equation}
	where the identical-particle factor for the two final-state axions has been included.
	
	In the limit $\Gaa\ll\Gphi$, the axion energy density produced by direct decay is fixed by the relative rate at which the inflaton transfers energy into axions and into the visible bath.  After reheating, the corresponding axion-to-visible radiation ratio is therefore
	\begin{equation}
		R_a^{\rm dec}
		=
		\frac{\Gaa}{\Gphi}
		=
		\frac{c_1^2}{128\pi}
		\frac{m_\phi^3}{\Lambda^2\Gphi}
		\equiv
		{\rm BR}(\phi\to aa).
		\label{eq:R_dec}
	\end{equation}
	Thus the direct-decay contribution to dark radiation is
	\begin{equation}
		\DNeff^{\rm dec}
		=
		\Arh\,{\rm BR}(\phi\to aa)\,,
		\label{eq:DNeff_dec}
	\end{equation}
	corresponding to the first term in Eq.~\eqref{eq:master_formula}.
	Using Eq.~\eqref{eq:Gamma_Trh}, one sees immediately that
	$\DNeff^{\rm dec}\propto\Trh^{-2}$, which decreases with reheating temperature.
	The reason is simple: a larger $\Trh$ corresponds to a larger visible
	reheating width, implying a smaller
	invisible branching fraction.
	
	\subsection{Upper bounds on inflaton invisible decays}
	
	Taking the representative Planck 2018 bound
	$\DNeff\leq0.34$ at 95\% C.L.~\cite{Planck:2018vyg}, and using
	$\Arh\simeq2.86$, one obtains
	\begin{equation}
		{\rm BR}(\phi\to aa)
		\lesssim
		1.2\times10^{-1}.
		\label{eq:BR_bound_current}
	\end{equation}
	Future measurements of $\DNeff$ would directly improve
	this branching-ratio bound.  Representative current and projected
	sensitivities are summarized in Table~\ref{tab:BR_bounds}.  For example, a
	CMB-HD-level sensitivity would reach
	\begin{equation}
		{\rm BR}(\phi\to aa)
		\lesssim
		9.4\times10^{-3}.
	\end{equation}
	We note that these bounds depend only on the final
	relativistic branching fraction, not on the microscopic form of the visible
	reheating operator. However, this interpretation assumes that the decay contribution dominates, which is valid in the low-reheating-temperature regime.

	\begin{table}[t]
		\centering
		\begin{tabular}{c c c}
			\hline\hline
			Reference & $\Delta N_{\rm eff}$ &
			${\rm BR}(\phi\to aa)_{\rm max}$ \\
			\hline
			Planck 2018 \cite{Planck:2018vyg} & $0.34$ & $1.2\times 10^{-1}$ \\
			BBN+CMB \cite{Yeh:2022heq} & $0.14$ & $4.9\times 10^{-2}$ \\
			SO \cite{SimonsObservatory:2018koc} & $0.10$ & $3.5\times 10^{-2}$ \\
			PICO \cite{NASAPICO:2019thw}  & $0.06$ & $2.1\times 10^{-2}$ \\
			CMB-HD \cite{CMB-HD:2022bsz} & $0.027$ & $9.4\times 10^{-3}$ \\
			\hline\hline
		\end{tabular}
		\caption{
			Representative upper bounds on the invisible relativistic branching fraction of the inflaton inferred from $\Delta N_{\rm eff}$.  The first two rows are current constraints, while the remaining rows are projected sensitivities.  The conversion uses $\DNeff^{\rm dec}=\Arh{\rm BR}(\phi\to aa)$ with $\Arh\simeq2.86$.
		}
		\label{tab:BR_bounds}
	\end{table}
	
	The same constraint can be written as a bound on the Wilson coefficient $c_1$.  From Eq.~\eqref{eq:Gamma_aa}, the direct-decay bound gives
	\begin{equation}
		|c_1|
		<
		\left[
		\frac{128\pi}{\Arh}
		\DNeff^{\rm max}
		\frac{\Lambda^2\Gphi}{m_\phi^3}
		\right]^{1/2}.
		\label{eq:c1_bound_Gamma}
	\end{equation}
	Using the reheating convention in Eq.~\eqref{eq:Gamma_Trh}, this becomes
	\begin{align}
		|c_1|
		&<
		\left[
		\frac{64\pi^2}{\Arh}
		\left(\frac{\grho(\Trh)}{10}\right)^{1/2}
		\DNeff^{\rm max}
		\frac{\Lambda^2\Trh^2}{\Mp m_\phi^3}
		\right]^{1/2}
		\nonumber\\
		&=
		0.16 \,
		\left(\frac{\DNeff^{\rm max}}{0.34}\right)^{1/2}
		\left(\frac{\grho(\Trh)}{106.75}\right)^{1/4}
		\left(\frac{2.86}{\Arh}\right)^{1/2}
		\left(\frac{10^{3} \,\Lambda}{\Mp}\right)
		\left(\frac{\Trh}{m_\phi}\right)
		\left(\frac{\Mp}{10^4\, m_\phi}\right)^{1/2}.
		\label{eq:c1_bound_Trh}
	\end{align}
	This form makes the parametric dependence transparent.  At fixed $\DNeff^{\rm max}$, the direct-decay bound becomes stronger for smaller $\Lambda$, lower $\Trh/m_\phi$, and larger $m_\phi$.  This is simply the statement that the inflaton decay width to axions scales as  $\Gaa\propto c_1^2/\Lambda^2$, while the visible reheating width grows as $\Gphi\propto \Trh^2$.  Thus a smaller cutoff strengthens the inflaton--axion coupling, whereas a larger visible reheating rate dilutes the axion branching fraction.
	
	For a typical high-scale inflaton mass $m_\phi\simeq 10^{13}\,{\rm GeV}$, perturbative reheating with $\Trh\lesssim m_\phi$, and a lower cutoff $\Lambda\lesssim10^{-4}\Mp$, the current $\DNeff$ bound corresponds to $c_1\lesssim{\cal O}(0.1)$.  If instead the kinetic coupling is Planck-suppressed, $\Lambda=M_P$, the same value of $c_1$ gives a much smaller partial width into axions, and the direct-decay constraint on $c_1$ is correspondingly weaker.  Conversely, at fixed $\Lambda$ and $m_\phi$, low reheating temperatures give the strongest direct-decay constraint because $\Gaa/\Gphi$ is larger.
	
	The direct-decay channel therefore plays two roles in the EFT map.  First, it provides the immediate connection to the invisible-branching-ratio bound on dark radiation.  Second, within the kinetic EFT it measures the linear response of the axion kinetic metric, namely the coefficient $c_1$ in $Z(\phi)$.  This makes it one axis of the full EFT map.  The second axis, discussed next, is the inflaton-annihilation channel, which probes the quadratic response of the same kinetic metric together with the exchange diagrams generated by the linear operator.
	
	\section{Annihilation}
	\label{sec:annihilation}
	
	The quadratic part of the kinetic function probes a different aspect of reheating.  Because the source is quadratic in $\rho_\phi$, this contribution depends on the density and duration of the reheating era.
	
	\subsection{Tree-level amplitude}
	
	The two-inflaton process $\phi\phi\to aa$ receives three tree-level contributions. The quadratic interaction in Eq.~\eqref{eq:interactions} gives a contact diagram. Two insertions of the linear interaction generate $t$- and $u$-channel axion exchange diagrams; see Fig.~\ref{fig:feynman_diagrams}.
	The corresponding nonrelativistic cross section is
	\begin{equation}
		\sigma_{\phi\phi\to aa}
		=
		\frac{(c_2-c_1^2)^2}{16\pi}
		\frac{m_\phi^2}{\Lambda^4}.
		\label{eq:sigmav}
	\end{equation}
	The event rate per unit volume is $\frac12n_\phi^2\sigma$, where $n_\phi=\rho_\phi/m_\phi$ is the number density of nonrelativistic inflatons. The factor of $1/2$ avoids double counting identical initial particles. Each annihilation event deposits energy $2m_\phi$ into axions. The axion energy-injection rate is therefore
	\begin{equation}
		Q_a^{\rm ann}
		= \rho_\phi (n_\phi \sigma) = 
		\frac{\rho_\phi^2}{m_\phi} \sigma_{\phi\phi\to aa}.
		\label{eq:Q_ann}
	\end{equation}
	Equivalently, the axion energy density obeys
	\begin{equation}
		\dot\rho_a^{\rm ann}+4H\rho_a^{\rm ann}
		=
		\frac{\rho_\phi^2}{m_\phi}
		\frac{(c_2-c_1^2)^2}{16\pi}
		\frac{m_\phi^2}{\Lambda^4}.
		\label{eq:rho_a_ann}
	\end{equation}
	The term $4H\rho_a^{\rm ann}$ accounts for the redshifting of relativistic axions after production. 
	
	\subsection{Reheating kernel}
	
	Eq.~\eqref{eq:rho_a_ann} is linear in $\rho_a^{\rm ann}$ and the source is proportional to $(c_2-c_1^2)^2$.  It is therefore useful to factor the final abundance into a microscopic coefficient and a cosmological kernel,
	\begin{equation}
		R_a^{\rm ann}
		=
		(c_2-c_1^2)^2\Kann .
		\label{eq:R_ann}
	\end{equation}
	The coefficient $(c_2-c_1^2)^2$ contains the particle-physics information from the nonrelativistic annihilation amplitude, while $\Kann$ contains the information about the energy scales and duration of the reheating.
	
	Eq.~\eqref{eq:rho_a_ann} can be conveniently solved by introducing  a comoving axion energy density $E \equiv \rho\, \mathfrak{a}^4$. This gives
	\begin{equation}
		\Kann
		=
		\frac{m_\phi}{16\pi\Lambda^4}
		\frac{1}{\rho_R(\af)}
		\int_{\ai}^{\af}
		\frac{d\mathfrak{a}}{\mathfrak{a}\,H(\mathfrak{a})}
		\left(\frac{\mathfrak{a}}{\af}\right)^4
		\rho_\phi^2(\mathfrak{a}).
		\label{eq:K_ann}
	\end{equation}
	The factor $\rho_\phi^2$ reflects the fact that two inflatons participate in the process.  The factor $(a/\af)^4$ redshifts the axion energy density produced at scale factor $a$ to the final time $\af$.  The lower limit $\ai$ denotes the onset of the controlled perturbative kinetic-EFT calculation (defined in Eq.~\eqref{eq:Phi_xi}), with $H(\ai)=H_i$.  The final scale factor $\af$ is chosen after visible reheating has completed.  At that stage both $\rho_a$ and $\rho_R$ redshift as radiation, so the ratio $\rho_a/\rho_R$ is independent of the precise choice of $\af$.
	
	During reheating, we have $H\propto a^{-3/2}$ and $\rho_\phi\propto a^{-3}$.  Using the reheating convention $H(\af)\simeq 2\Gphi/3$, Eq.~\eqref{eq:K_ann} becomes\footnote{See Appendix \ref{app:kernel} for  more details.}
	\begin{align}
		\Kann^{\rm ana}
		&=
		\frac{1}{4\pi}
		\frac{\Mp^2m_\phi\Gphi}{\Lambda^4}
		\left[
		\left(\frac{a_f}{a_i}\right)^{1/2}
		-1
		\right]
		\nonumber\\
		&=
		\frac{1}{4\pi}
		\frac{\Mp^2m_\phi\Gphi}{\Lambda^4}
		\left[
		\left(\frac{3H_i}{2\Gphi}\right)^{1/3}
		-1
		\right].
		\label{eq:K_ann_ana}
	\end{align}
	The first factor shows the microscopic suppression $\Lambda^{-4}$ of the annihilation process.  The bracket measures the time available for inflaton annihilation between the controlled onset and the end of reheating.

	The annihilation contribution to dark radiation can be written
	\begin{equation}
		\DNeff^{\rm ann}
		=
		\Arh(c_2-c_1^2)^2\Kann\,,
		\label{eq:DNeff_ann}
	\end{equation}
	corresponding to the second term in Eq.~\eqref{eq:master_formula}.
	Its dependence on the reheating temperature is qualitatively different from the direct-decay contribution.  When the controlled inflaton-oscillation era is long, $H_i\gg\Gphi$, the first term in the bracket of Eq.~\eqref{eq:K_ann_ana} dominates, giving $\DNeff^{\rm ann}\propto\Trh^{4/3}$.  Direct decay behaves in the opposite way, $\DNeff^{\rm dec}\propto\Trh^{-2}$.
	
	The reason is physical.  Direct decay is a branching-ratio probe.  Increasing $\Trh$ increases the visible reheating width $\Gphi$, so a fixed partial width into axions represents a smaller fraction of the total inflaton energy loss.  Annihilation is different: it is sourced by two powers of the inflaton density.  A larger $\Trh$ means that reheating completes earlier, while the oscillating inflaton background is still denser.  In the regime $H_i\gg\Gphi$, this density enhancement wins over the shorter available time, and the net annihilation contribution grows with $\Trh$.  The growth stops only when the interval between the controlled EFT onset and the end of reheating becomes short.
	
	Thus direct decay and inflaton annihilation are not two descriptions of the same physics.  They probe different properties of reheating.  Direct decay measures an invisible branching fraction and is strongest at low $\Trh$, where the visible width is small.  Inflaton annihilation instead measures the density history of the oscillating inflaton background and can become competitive at higher $\Trh$.  The crossing of these contributions is the phenomenologically important regime missed by decay-only treatments based solely on an invisible branching fraction.   In this regime, $\DNeff$ is no longer a one-parameter constraint, but a probe of the inflaton dependence of the axion kinetic metric in two independent directions.
	
	A bound on $\DNeff$ gives a bound on the invariant annihilation combination
	\begin{equation}
		|\cann|
		<
		\left[
		\frac{\DNeff^{\rm max}}{\Arh\Kann}
		\right]^{1/2},
		\qquad
		\cann = c_2-c_1^2 .
		\label{eq:cann_bound}
	\end{equation}
	This is the analogue of the direct-decay bound on $c_1$: the decay channel constrains the linear response of the kinetic metric, while the annihilation channel constrains the distance from the blind line $c_2=c_1^2$.
	
	To make the parametric dependence explicit, define the dimensionless duration factor
	\begin{equation}
		\mathcal B_{\rm ann}
		\equiv
		\left(\frac{3H_i}{2\Gphi}\right)^{1/3}-1 .
		\label{eq:Bann_def}
	\end{equation}
	The perturbative annihilation interval exists only for $B_{\rm ann}>0$, equivalently $H_i>2\Gphi/3$.  If this condition is not satisfied, visible reheating has completed before the controlled perturbative EFT regime begins, and there is no remaining inflaton-annihilation interval to compute.
	
	Using Eq.~\eqref{eq:K_ann_ana} together with the reheating convention in Eq.~\eqref{eq:Gamma_Trh}, Eq.~\eqref{eq:cann_bound} becomes
	\begin{align}
		|\cann|
		&<
		\left[
		\frac{8}{\Arh}
		\left(\frac{10}{\grho(\Trh)}\right)^{1/2}
		\DNeff^{\rm max}
		\frac{\Lambda^4}{\Mp m_\phi \Trh^2}
		\frac{1}{\mathcal B_{\rm ann}}
		\right]^{1/2}
		\nonumber\\
		&=
		0.54 
		\left(\frac{\DNeff^{\rm max}}{0.34}\right)^{1/2}
		\left(\frac{2.86}{\Arh}\right)^{1/2}
		\left(\frac{106.75}{\grho(\Trh)}\right)^{1/4}
		\left(\frac{\Lambda}{\Mp}\right)^2
		\left(\frac{m_\phi}{\Trh}\right)
		\left(\frac{\Mp}{m_\phi}\right)^{3/2}
		\left(\frac{10}{\mathcal B_{\rm ann}} \right)^{1/2}
		\label{eq:cann_bound_Trh}
	\end{align}
	This form should be compared with the direct-decay bound in Eq.~\eqref{eq:c1_bound_Trh}.  The annihilation bound scales as $\Lambda^2$, whereas the decay bound scales as $\Lambda$.  This difference reflects the fact that the annihilation rate is suppressed by $\Lambda^{-4}$, while the decay width is suppressed by $\Lambda^{-2}$.  The annihilation bound also depends on the available duration of the controlled reheating  era through $\mathcal B_{\rm ann}$.  As $H_i$ approaches $2\Gphi/3$, $\mathcal B_{\rm ann}$ tends to zero and the bound becomes weak, because the inflaton disappears before perturbative annihilation has time to operate.
	
	The duration factor is not an independent parameter.  In the controlled
	lower-cutoff regime we initialize the perturbative kinetic EFT at
	\begin{equation}
		\Phi_i=\xi\Lambda,
		\qquad
		H_i=\frac{\xi m_\phi\Lambda}{\sqrt6\Mp},
		\label{eq:Hi_xi_ann}
	\end{equation}
	so that
	\begin{equation}
		\mathcal B_{\rm ann}
		=
		\left[
		\frac{3\xi m_\phi\Lambda}{2\sqrt6\,\Mp\Gphi}
		\right]^{1/3}
		-1 .
		\label{eq:Bann_xi}
	\end{equation}
	To derive $H_i$ in Eq.~\eqref{eq:Hi_xi_ann}, we have utilized $ \rho_\phi = m_\phi^2\Phi_i^2/2 =  3H_i^2 M_P^2$. 
	Thus the annihilation bound is controlled by two competing effects.  A
	smaller cutoff enhances the microscopic annihilation rate through the
	$\Lambda^{-4}$ dependence of the cross section.  At the same time, EFT
	control requires the perturbative calculation to start only when the
	oscillation amplitude has redshifted to the controlled range
	$\Phi_i\simeq\xi\Lambda$, which reduces the available reheating history.
	The factor $\mathcal B_{\rm ann}$ precisely measures this remaining
	duration.
	
	This is the main difference from the decay channel.  Direct decay is
	controlled by the branching ratio of a single inflaton quantum and therefore
	constrains $c_1$.  Inflaton annihilation is controlled by both the
	two-inflaton amplitude and the lifetime of the inflaton, and
	therefore constrains the distance from the blind line, $c_2=c_1^2$.
	In this sense $\DNeff$ probes two different derivatives of the same axion
	kinetic metric: the linear response through decay, and the coherent
	quadratic response through annihilation.
	
	\begin{figure}[!h]
		\def\sepf{0.6}
		\centering
		\includegraphics[scale=\sepf]{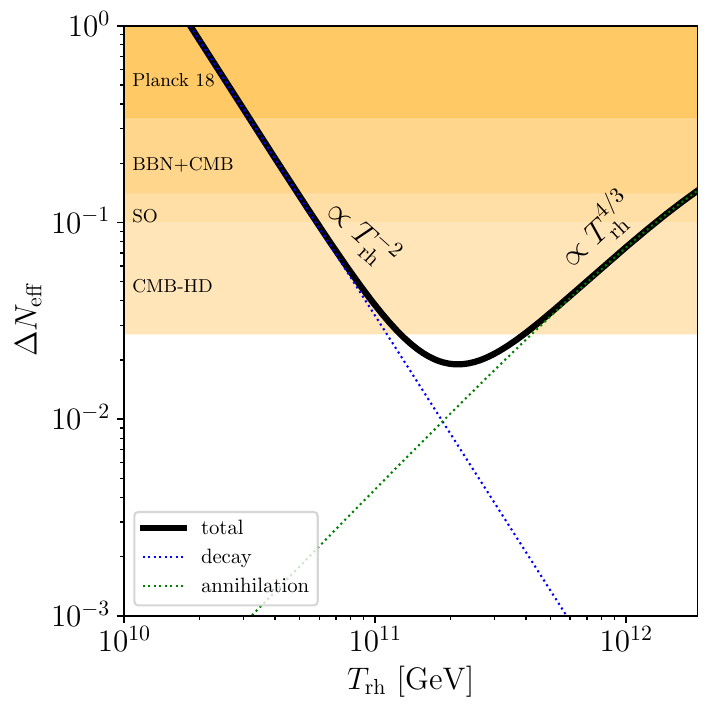}
		\caption{
			$\DNeff$ as a function of the reheating temperature for the benchmark
			$c_1=10^{-3}$, $c_2=0.3$, $\Lambda=10^{14}\,{\rm GeV}$, and $\xi=0.5$.
			The blue dotted, green dotted, and black solid curves denote the direct-decay,
			inflaton-annihilation, and total contributions, respectively.  The decay contribution
			falls as $\Trh^{-2}$, while the annihilation contribution grows approximately as
			$\Trh^{4/3}$.  The shaded bands indicate representative
			current and projected sensitivities to $\DNeff$.
		}
		\label{fig:DNeff_Trh}
	\end{figure}
	\section{Results}
	\label{sec:map}
	
	We now combine the direct-decay and inflaton-annihilation contributions into the full EFT map. 
	The benchmarks are chosen to illustrate the transition between the decay-dominated and annihilation-dominated regimes while remaining within the controlled kinetic-EFT expansion. 
	For the benchmarks shown below, the kinetic expansion remains perturbative in the controlled regime. 
	In particular, the largest values used satisfy $|c_1|\xi < 1$ and $|c_2|\xi^2/2 < 1$, so the modulation of the axion kinetic metric remains small. 
	Throughout this section we take a representative high-scale inflaton mass  $m_\phi \simeq 10^{13}\,{\rm GeV}$.
	
	Figure~\ref{fig:DNeff_Trh} illustrates the basic mechanism behind the EFT map.  The shaded bands show the representative current and projected sensitivities summarized in Table~\ref{tab:BR_bounds}. The two production channels have opposite dependence on the reheating temperature.  Direct decay is controlled by an invisible branching ratio: increasing the visible reheating width reduces the fraction of inflaton energy deposited into axions, giving $\DNeff^{\rm dec}\propto\Trh^{-2}$.  Inflaton annihilation is instead controlled by the density history of the oscillating inflaton background.  In the matter-dominated regime, its contribution grows approximately as $\DNeff^{\rm ann}\propto\Trh^{4/3}$, because larger $\Trh$ corresponds to reheating while the inflaton energy density is still larger.
	
	The minimum of the total curve marks the transition between the decay-dominated and annihilation-dominated regimes.  This crossing region is not captured by a decay-only description in terms of an invisible branching fraction, because $\DNeff$ is then sensitive both to the linear coefficient $c_1$ and to the annihilation combination $c_2-c_1^2$. 
	
	The bending of the total curve is the observable consequence of treating the kinetic EFT consistently.  A decay-only analysis would suggest that the dark-radiation constraint is strongest at low $\Trh$, whereas an annihilation-only analysis would suggest the opposite.  When both operators are kept, these two limits are joined into a finite allowed band: low reheating temperatures are constrained by direct decay, while high reheating temperatures are constrained by inflaton annihilation.  For the benchmark in Fig.~\ref{fig:DNeff_Trh}, an SO-level sensitivity would be able to bound the reheating temperature to be  $5\times10^{10}\ {\rm GeV}
	\lesssim
	\Trh
	\lesssim
	10^{12}\ {\rm GeV}.$
	This illustrates one of the central phenomenological points of the map: $\Delta N_{\rm eff}$ can constrain the reheating temperature through the competition of two microscopic production channels, rather than only placing a one-sided bound on an invisible branching fraction.
	
	\begin{figure}[!t]
		\def\sepf{0.6}
		\centering
		\includegraphics[scale=\sepf]{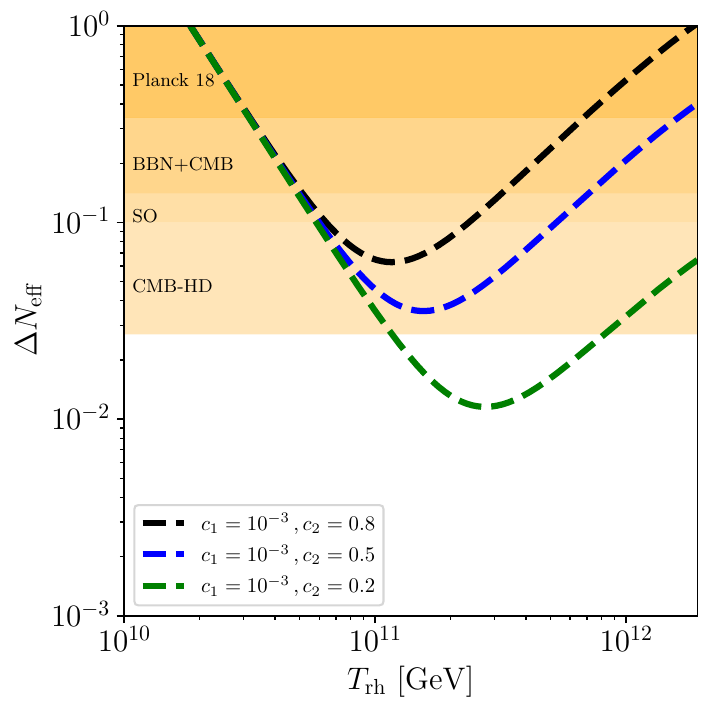}
		\includegraphics[scale=\sepf]{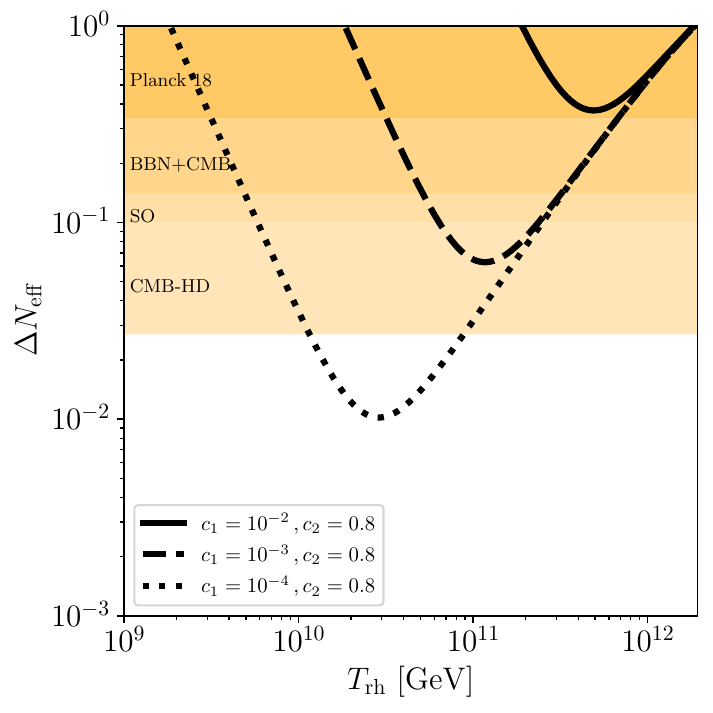}
		\caption{
			Dependence of $\DNeff$ on the EFT coefficients for
			$\Lambda=10^{14}\,{\rm GeV}$ and $\xi=0.5$.
			Left: $c_1=10^{-3}$ is fixed while $c_2$ is varied.  The curves coincide
			at low $\Trh$, where direct decay dominates, and separate at high $\Trh$,
			where inflaton annihilation becomes important.
			Right: $c_2=0.8$ is fixed while $c_1$ is varied.  The curves separate at low
			$\Trh$ because the decay contribution scales as $c_1^2$, but merge at high
			$\Trh$, where the signal is controlled by inflaton annihilation.
		}
		\label{fig:DNeff_Trh_2}
	\end{figure}
	
	Figure~\ref{fig:DNeff_Trh_2} shows how this structure depends on the EFT coefficients.  In the left panel, $c_1$ is fixed and $c_2$ is varied.  Since the annihilation contribution is proportional to $(c_2-c_1^2)^2$, decreasing $c_2$ suppresses the high-$\Trh$ branch.  At low $\Trh$, inflaton annihilation is negligible and all curves approach the same decay-dominated behavior.  The position of the minimum therefore tracks the point at which inflaton annihilation becomes competitive with direct decay.
	
	In the right panel, $c_2$ is fixed and $c_1$ is varied.  This changes the direct-decay contribution, which scales as $c_1^2$, while leaving the high-$\Trh$ annihilation branch almost unchanged for the benchmarks shown, where $c_2\gg c_1^2$.  The curves therefore separate at low $\Trh$, where direct decay dominates, but merge at high $\Trh$, where the signal is controlled by inflaton annihilation.
	
	Together, Figs.~\ref{fig:DNeff_Trh} and~\ref{fig:DNeff_Trh_2} show why $\DNeff$ is naturally a two-dimensional probe of the axion kinetic metric.  The coefficient $c_1$ controls the branching-ratio direction, while $c_2-c_1^2$ controls the inflaton-annihilation direction.  We now display this structure directly in the $(c_1,c_2)$ plane.
	
	\begin{figure}[!ht]
		\def\sepf{0.6}
		\centering
		\includegraphics[scale=\sepf]{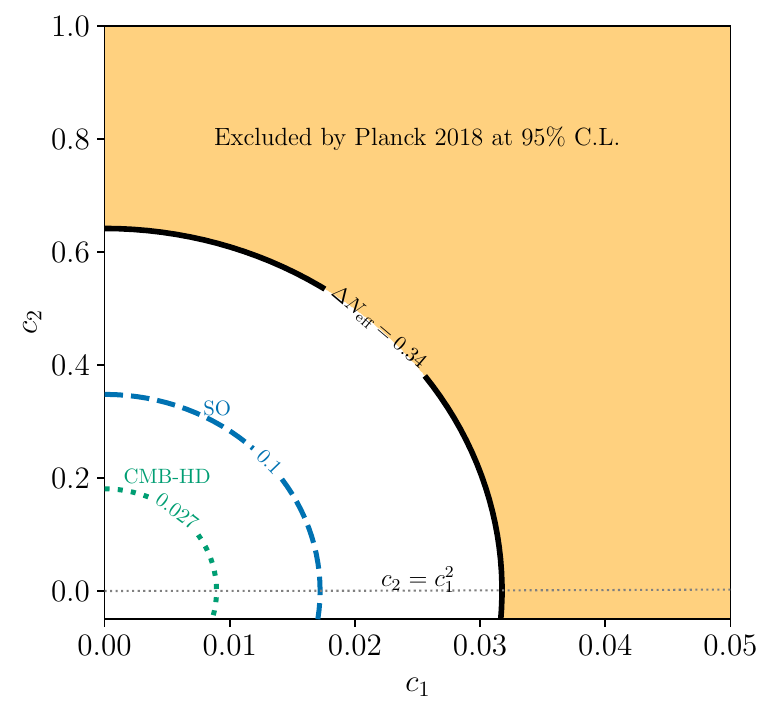}
		\caption{
			Two-dimensional EFT map in the $(c_1,c_2)$ plane for
			$\Lambda=10^{14}\,{\rm GeV}$, $\xi=0.5$, and
			$\Trh=10^{12}\,{\rm GeV}$.  The orange region is excluded by the
			Planck 2018 bound $\DNeff>0.34$ at 95\% C.L.  The black, blue dashed,
			and green dotted curves show the reference contours
			$\DNeff=0.34$, $0.10$, and $0.027$, corresponding to the Planck 2018 bound
			and representative SO and CMB-HD sensitivities.  The gray dotted curve
			denotes $c_2=c_1^2$.
		}
		\label{fig:c1_c2_map}
	\end{figure}
	
	Figure~\ref{fig:c1_c2_map} is the central EFT map.  The excluded region is curved rather than aligned with either axis because the two contributions depend on different combinations of Wilson coefficients.  Direct decay depends on $c_1^2$, while inflaton annihilation depends on $(c_2-c_1^2)^2$.  The line $c_2=c_1^2$ is therefore a blind line for the nonrelativistic inflaton-annihilation amplitude: along it, the contact and exchange contributions cancel in the annihilation channel.
	
	The shape of the contours has a simple interpretation.  Moving in the $c_1$ direction increases the direct-decay contribution.  Moving away from $c_2=c_1^2$ increases the inflaton-annihilation contribution.  The current Planck bound already excludes part of the large-coupling region, while future sensitivities would probe closer to the origin.  This is the main result of the map: dark radiation constrains both the linear response and the quadratic annihilation response of the axion kinetic metric.
	
	\begin{figure}[!ht]
		\def\sepf{0.5}
		\centering
		\includegraphics[scale=\sepf]{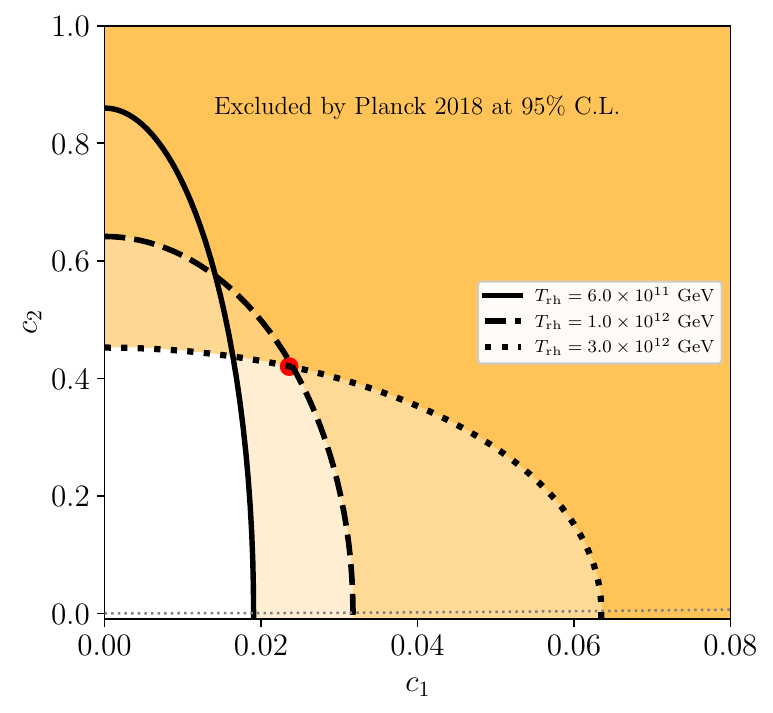}
		\includegraphics[scale=\sepf]{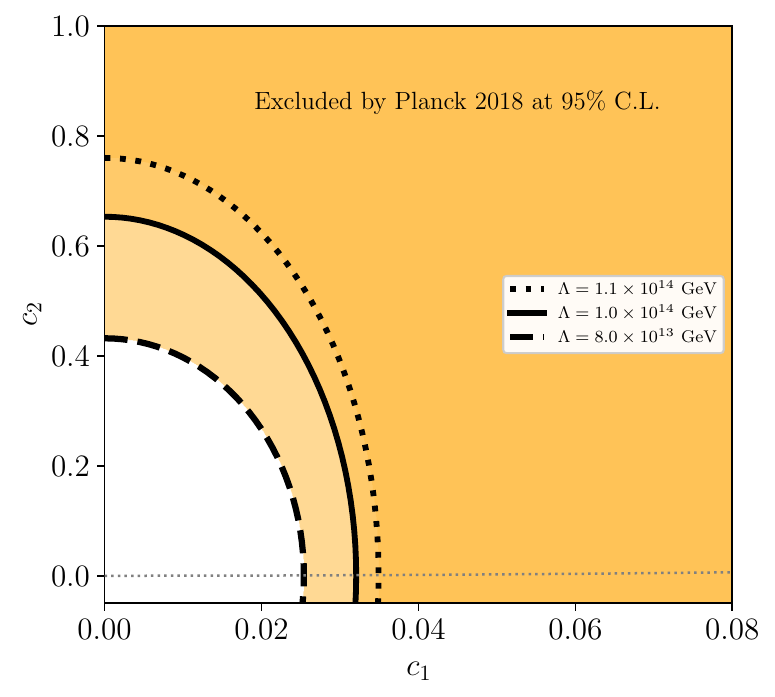}
		\includegraphics[scale=\sepf]{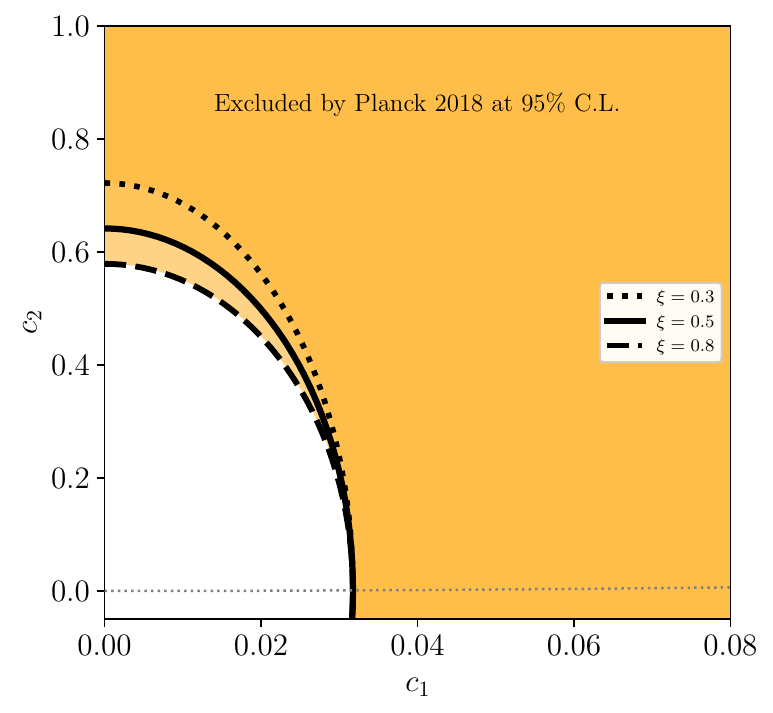}
		\caption{
			Dependence of the Planck 2018 exclusion region on the cosmological and EFT
			parameters.  The contours show $\DNeff=0.34$ in the $(c_1,c_2)$ plane, and
			the shaded regions are excluded by $\DNeff>0.34$ at 95\% C.L.
			Upper left: variation with $\Trh$ at fixed $\Lambda=10^{14}\,{\rm GeV}$ and
			$\xi=0.5$.  Upper right: variation with $\Lambda$ at fixed
			$\Trh=10^{12}\,{\rm GeV}$ and $\xi=0.5$.  Lower panel: variation with $\xi$
			at fixed $\Lambda=10^{14}\,{\rm GeV}$ and $\Trh=10^{12}\,{\rm GeV}$.
			The gray dotted curve denotes $c_2=c_1^2$, where the
			inflaton-annihilation amplitude cancels.
		}
		\label{fig:c1_c2_map_2}
	\end{figure}
	
	Figure~\ref{fig:c1_c2_map_2} shows how the Planck boundary moves when the reheating temperature, the EFT scale, and the controlled-onset parameter are varied.  Each parameter changes a different part of the production history.
	
	In the upper-left panel, increasing $\Trh$ moves the Planck contour toward larger $c_1$ but smaller $c_2$.  This follows from the opposite $\Trh$ dependence of the two production channels.  A larger $\Trh$ corresponds to a larger visible reheating width, which suppresses the direct-decay branching fraction and therefore permits a larger value of $c_1$ at fixed $\DNeff$.  At the same time, reheating completes earlier, when the oscillating inflaton background is denser.  This enhances inflaton annihilation and therefore requires a smaller value of $c_2-c_1^2$ along the same $\DNeff$ contour.
	
	For fixed Wilson coefficients, the crossing of the Planck contours can translate the $\DNeff$ bound into a finite window for the reheating temperature as in Figs.~\ref{fig:DNeff_Trh} and \ref{fig:DNeff_Trh_2}.  For example, for the benchmark point marked by the red dot in the upper-left panel of Figure~\ref{fig:c1_c2_map_2}, $(c_1,c_2)=\left(1.6\times10^{-2},\,4.4\times10^{-1}\right),$
	the Planck 2018 bound implies
	\begin{align}
		6.0\times10^{11}\ {\rm GeV}
		\lesssim
		\Trh
		\lesssim
		3.0\times10^{12}\ {\rm GeV}.
		\label{eq:Trh_window}
	\end{align}

	The upper-right panel shows the dependence on the EFT scale $\Lambda$.  Increasing $\Lambda$ weakens the inflaton--axion kinetic interaction, so the same dark-radiation abundance requires larger Wilson coefficients.  The effect is especially visible in the annihilation direction, because inflaton annihilation is suppressed by $\Lambda^{-4}$, whereas direct decay is suppressed by $\Lambda^{-2}$.  The outward motion of the contours with increasing $\Lambda$ is therefore the expected EFT scaling.
	
	The lower panel shows the dependence on $\xi$, which fixes the onset of the controlled kinetic-EFT calculation through $\Phi_i=\xi\Lambda$.  Increasing $\xi$ initializes the perturbative calculation at a larger inflaton amplitude and therefore includes a longer period of inflaton annihilation.  This mainly affects the $c_2$ direction, while the direct-decay direction is much less sensitive to $\xi$.  Consequently, larger $\xi$ allows a smaller annihilation coefficient to produce the same $\DNeff$.
	
	These panels make clear that the Planck constraint is not a one-dimensional bound on an invisible branching fraction.  Its position and shape depend on the competition between direct decay, inflaton annihilation, the duration of the controlled inflaton-oscillation era, and the EFT scale.  The natural result is therefore a two-dimensional map of the kinetic function, rather than a single limit on $c_1$.
	
	\section{Conclusions}
	\label{sec:conclusions}
	
	We have developed a perturbative EFT map from an inflaton-dependent axion kinetic metric to axion dark radiation.  The organizing principle is the axion shift symmetry: at leading order, the inflaton dependence enters through the kinetic function $Z(\phi)$.  Expanding this function near the minimum of the inflaton potential gives a linear response and a quadratic response, which correspond physically to two different sources of axion radiation.  The EFT framework is useful because it treats these sources in a common language, rather than describing dark radiation only through an invisible inflaton branching fraction.
	
	The central result is Eq.~\eqref{eq:master_formula}.  The late-time dark-radiation abundance receives a direct-decay contribution controlled by the linear response of $Z(\phi)$ and an inflaton-annihilation contribution controlled by the quadratic response.  The latter must be computed within the same EFT expansion: the contact diagram from $\phi^2(\partial a)^2$ and the exchange diagrams generated by two insertions of $\phi(\partial a)^2$ contribute to the same nonrelativistic annihilation amplitude.  This fixes the appropriate annihilation coefficient to be $\cann=c_2-c_1^2$, but the main consequence is broader: $\Delta N_{\rm eff}$ probes more than one direction in the kinetic function.
	
	The main phenomenological result is the opposite reheating-temperature dependence of the two channels.  Direct decay is an invisible-branching-ratio probe, as shown in Eq.~\eqref{eq:DNeff_dec} and Table~\ref{tab:BR_bounds}, and scales as $\DNeff^{\rm dec}\propto \Trh^{-2}$.  Inflaton annihilation is a reheating-history probe and grows approximately as $\DNeff^{\rm ann}\propto \Trh^{4/3}$ during  reheating regime.  This behavior is shown in Figs.~\ref{fig:DNeff_Trh} and~\ref{fig:DNeff_Trh_2}: the total signal bends because the decay-dominated and annihilation-dominated regimes are connected within the same EFT.
	
	This bending is the qualitative feature missed by incomplete treatments.  A decay-only analysis gives a one-sided conclusion, with the strongest constraint at low $\Trh$.  An annihilation-only analysis gives the opposite one-sided conclusion, with the strongest constraint at high $\Trh$.  The full kinetic EFT   can instead give a finite reheating-temperature window for fixed microscopic parameters: too low a reheating temperature is constrained by direct decay, while too high a reheating temperature is constrained by inflaton annihilation.  Thus the EFT organization changes the interpretation of $\Delta N_{\rm eff}$ from a single branching-ratio bound into a probe of the reheating window; see e.g. Eq.~\eqref{eq:Trh_window}.
	
	The two-dimensional maps in Figs.~\ref{fig:c1_c2_map} and~\ref{fig:c1_c2_map_2} display this structure in the Wilson-coefficient plane.  The exclusion and sensitivity contours are not aligned with either axis, because changing $c_1$ changes the decay direction, while changing the quadratic response changes the annihilation direction.  Varying $\Trh$, the cutoff $\Lambda$, and the controlled-onset parameter $\xi$ moves the contours in physically distinct ways: $\Trh$ controls the competition between decay and annihilation, $\Lambda$ controls the strength of the kinetic interaction, and $\xi$ fixes the onset of the perturbative kinetic-EFT calculation, or equivalently the scale $H_i$ from which the controlled annihilation history begins.
	
	In summary, axion dark radiation from reheating is not only a constraint on how often the inflaton decays invisibly.  In an inflaton-dependent kinetic EFT, $\Delta N_{\rm eff}$ probes both the microscopic couplings and the cosmological history of the inflaton.  Treating direct decay and inflaton annihilation consistently leads to phenomenological features -- including a possible lower and upper bound on $\Trh$ -- that are absent in decay-only or annihilation-only descriptions.
	\begin{acknowledgments}
		The author is grateful to  Jim Cline  for  useful discussions. 
		This work was supported by the Natural Sciences and Engineering Research Council (NSERC) of Canada.
	\end{acknowledgments}
	
	\appendix

	\section{Analytic annihilation kernel}
	\label{app:kernel}
	
	The kernel in Eq.~\eqref{eq:K_ann} is the quantity used for the EFT map.  This appendix derives the analytic  approximation in Eq.~\eqref{eq:K_ann_ana}.  The lower limit of the integral is the onset of the perturbative kinetic-EFT description, not necessarily the end of inflation.  This distinction matters when $\Lambda\ll\Mp$.  If the kinetic expansion is only controlled after the oscillation amplitude has redshifted to $\Phi_i=\xi\Lambda$, then the analytic kernel should be evaluated with $H_i=m_\phi\Phi_i/(\sqrt6\Mp)$.  The derivation below assumes that this onset occurs before the end of reheating, $H_i>H_{\rm rh}$.
	
	During  reheating, we have
	\begin{equation}
		\rho_\phi\simeq3\Mp^2H^2,
		\qquad
		H\propto a^{-3/2}.
	\end{equation}
	With the reheating convention used in this paper, the end of reheating is estimated by $H_{\rm rh}=2\Gphi/3$. Therefore
	\begin{equation}
		\frac{\mathfrak{a}}{\arh}
		=
		\left(\frac{H_{\rm rh}}{H}\right)^{2/3}
		=
		\left(\frac{2\Gphi}{3H}\right)^{2/3}.
	\end{equation}
	The annihilation-produced axion energy density at $\arh$ is obtained by integrating the source in Eq.~\eqref{eq:Q_ann} and redshifting each contribution from its production time to $\arh$,
	\begin{equation}
		\rho_\mathfrak{a}^{\rm ann}(\arh)
		=
		\int_{\ai}^{\arh}
		\frac{da}{aH}
		\left(\frac{\mathfrak{a}}{\arh}\right)^4
		\frac{\rho_\phi^2}{m_\phi} \sigma_{\phi\phi\to aa}.
	\end{equation}
	Using $d\ln \mathfrak{a}=-(2/3)d\ln H$, this becomes
	\begin{align}
		\rho_a^{\rm ann}(\arh)
		&=
		\frac{2}{3}
		\int_{H_{\rm rh}}^{H_i}
		\frac{dH}{H^2}
		\left(\frac{H_{\rm rh}}{H}\right)^{8/3}
		\frac{9\Mp^4H^4}{m_\phi}  \sigma_{\phi\phi\to aa}
		\nonumber\\
		&=
		18
		\frac{\Mp^4}{m_\phi}
		H_{\rm rh}^{8/3}
		\left(H_i^{1/3}-H_{\rm rh}^{1/3}\right)   \sigma_{\phi\phi\to aa}.
	\end{align}
	At reheating, the radiation density is approximated by $\rho_R(\arh)\simeq 3\Mp^2H_{\rm rh}^2$. Dividing by this quantity gives
	\begin{align}
		R_a^{\rm ann}
		\simeq&
		6
		\frac{\Mp^2H_{\rm rh}}{m_\phi}
		\left[
		\left(\frac{H_i}{H_{\rm rh}}\right)^{1/3}-1
		\right]    \sigma_{\phi\phi\to aa} \nonumber \\
		& \simeq
		(c_2-c_1^2)^2
		\frac{1}{4\pi}
		\frac{\Mp^2m_\phi\Gphi}{\Lambda^4}
		\left[
		\left(\frac{3H_i}{2\Gphi}\right)^{1/3}-1
		\right].
	\end{align}
	where $H_{\rm rh}=2\Gphi/3$ and Eq.~\eqref{eq:sigmav}  have been used.
	This identifies the analytic kernel in Eq.~\eqref{eq:K_ann_ana}.
	\bibliographystyle{JHEP}
	\bibliography{biblio}
	
\end{document}